\documentclass[%
superscriptaddress,
nofootinbib,
amsmath,amssymb,
aps,
pra,
longbibliography,
showkeys,
notitlepage
]{revtex4-1}

\usepackage[left=2.5cm,top=3cm,right=2.5cm,bottom=3cm,bindingoffset=0.5cm]{geometry}
\usepackage{graphicx}
\usepackage{hyperref}
\usepackage[bitstream-charter, greekfamily=default]{mathdesign}
\usepackage{lipsum}
\usepackage{xcolor}
\usepackage{subeqnarray}

\usepackage{epstopdf, epsfig}

\usepackage{natbib}

\usepackage{amssymb}
\usepackage{amsmath,amsfonts}

\usepackage{ulem}
 \usepackage{color}

\newcommand{\bnabla}{\boldsymbol{\nabla}}
\newcommand{\bcdot}{\boldsymbol{\cdot}}

\renewcommand{\epsilon}{\varepsilon}

\begin{document}

\title{Unsteady Wave Drag on a Disturbance Moving Along an Arbitrary Trajectory} 

\author{Lucas Gierczak-Galle}
\affiliation{{E}cole Normale Sup\'erieure, 45 rue d'Ulm, 75005 Paris, France}
\affiliation{UMR CNRS 7083 Gulliver, ESPCI Paris, PSL Research University, 75005 Paris, France}

\author{Assil Fadle}%
\affiliation{{E}cole Normale Sup\'erieure, 45 rue d'Ulm, 75005 Paris, France}
\affiliation{UMR CNRS  7646 LadHyX, {E}cole polytechnique, 91128 Palaiseau Cedex, France\medskip}

\author{Maxence Arutkin}%
\affiliation{UMR CNRS 7083 Gulliver, ESPCI Paris, PSL Research University, 75005 Paris, France}

\author{{E}lie Rapha\"el}
\affiliation{UMR CNRS 7083 Gulliver, ESPCI Paris, PSL Research University, 75005 Paris, France}

\author{Michael Benzaquen}%
\affiliation{UMR CNRS  7646 LadHyX, {E}cole polytechnique, 91128 Palaiseau Cedex, France\medskip}

\date{\today}

\begin{abstract}
We derive analytical formulas for the wake and wave drag of a disturbance moving arbitrarily at the air-water interface. We show that, provided a constant velocity is reached in finite time, the unsteady surface displacement converges to its well-known steady counterpart as given by Havelock's famous formula. Finally we assess, in a specific situation, to which extent one can rightfully use Havelock's steady wave drag formula for non-uniform motion  (quasi-static). Such an approach can be used to legitimize or discredit a number of studies which used steady wave drag formulas in unsteady situations.
\end{abstract}

\maketitle
	
	
	\section{Introduction}
		Water waves have fascinated physicists and mathematicians for centuries.  Among them, Lagrange was the first to derive the governing equations~\cite{darrigol2005worlds,lighthill1978waves} and Kelvin to  account successfully for the famous V-shape pattern of the wake behind a ship using stationary phase arguments~\cite{kelvin1887ship}. Kelvin's theory was recently brought up to date thanks to airborne observations of ship wakes~\cite{rabaud2013ship,darmon2014kelvin}, revealing more intricate phenomena which would not have been observable back in the day.\medskip
		
		The wake of a moving disturbance, say, a ship, naturally carries energy radiated by the source. Such energy loss translates into a drag force exerted on the disturbance and opposing its motion, commonly called \textit{wave drag} or \textit{wave resistance}~\cite{wehausen1960surface}. Both Havelock~\cite{havelock1932theory} and  Michell~\cite{michell1898xi} proposed methods to compute the wave drag of a body moving steadily at the air-water interface, which have been extensively used by both physicists and naval engineers in the shipbuilding industry~\cite{lighthill1978waves,parnell2001wakes} throughout the past century. Numerous experimental and theoretical studies have focused on the extension of their results to account for the aspect ratio  of the body~\cite{benzaquen2014wake,boucher2018thin}, its front-back asymmetry \cite{benham2019wave}, or varying  depth~\cite{benham2020hysteretic} with applications for hull design and rowing sports to name a few.
		At smaller scales, where capillarity is no longer negligible, the analysis of the wave drag is also relevant~\cite{raphael1996capillary}, notably to understand 
		the biolocomotion of certain insects and beetles~\cite{voise2010management,hu2003hydrodynamics,buhler2007impulsive}. The highly unsteady nature of the propulsion mechanisms of such insects revealed the importance of being able to properly account for unsteady effects in the wave drag~\cite{steinmann2018unsteady}.\medskip
		
		Lacking a general formula to compute the wave drag for  unsteady motion, a number of studies have used the steady Havelock formula as if it were applicable, see e.g.~\cite{le2011wave,benham2020hysteretic}. While in some cases (likely quasi-static) this might be justified, one may rightfully argue that it will lead to inaccurate conclusions in others. 
		\medskip
		
		In the present paper, we extend Havelock's theory to compute the unsteady wake and wave drag of disturbance with given trajectory $\mathbf r_0(t) = (x_0(t),y_0(t))$, which has no other constraints than being smooth enough, typically $\mathcal{C}^1$, see below. We obtain a general formula for both pure gravity, and capillary-gravity waves allowing to compute the wake and wave drag for any $\mathbf r_0(t)$. We illustrate our results for uniformly accelerated motion, which allows to determine an acceleration threshold below which Havelock's steady wave drag formula can be considered accurate to some extent. 	
	{See also the interesting study by Dutykh \& Dias~\cite{dutykh2007water} for an analysis of the unsteady waves generated by a moving bottom.}
    

	\section{Unsteady wakes}\label{sec:wakes}
			In this section we derive the surface elevation caused by a moving disturbance at the air-water interface. We assume   irrotational flow of an inviscid and infinitely deep fluid of constant and uniform density $\rho$, extending infinitely in the $(x,y)$ plane. The surface elevation of the fluid is denoted $\zeta(\mathbf r, t)$ with $\mathbf{r} = (x, y)$. Following Havelock's method, the moving disturbance 
			is modeled by an external pressure field $P_\text{ext}(\mathbf{r} , t)$ applied to the fluid surface, on top of the atmospheric pressure $P_{\mathrm{atm}}$. \medskip
			
			Hereafter we shall restrict to a linearized setting in which the waves amplitude always remain small compared to the wavelength, allowing to neglect second-order terms and make extensive use of Fourier transform. The disturbance trajectory is given by $\mathbf{r_0}(t)$, assumed to be  smooth enough (typically $\mathcal{C}^1$), such that:
			\begin{equation}
				P_\text{ext}(\mathbf{r}, t) = P_{\text{e}}(\mathbf{r} - \mathbf{r_0}(t))\, ,
			\end{equation}
			with $P_{\text{e}}(\mathbf{r})$ the pressure field at $t = 0$ at position $\mathbf{r}$. Denoting by ~$\hat{}$~ the two-dimensional Fourier Transform with respect to $\mathbf r$, it follows that:
			\begin{equation}
			   \hat P_\text{ext}(\mathbf{k}, t) = \hat P_{\text{e}}(\mathbf{k}) \text{e}^{-\text{i} \mathbf{k} \bcdot \mathbf{r_0}(t)}. 
			   \label{translation}
			\end{equation}	
			Denoting $\mathbf u$ the velocity field in the fluid, the linearized Euler equations read:
			\begin{subeqnarray}\label{euler_equation}
				\partial_t \mathbf{u}  &=& - \bnabla P/\rho + \mathbf{g} \, ,\\
				\bnabla \bcdot \mathbf{u} &=&0 \, ,\slabel{eq:incomp}
			\end{subeqnarray}
			where $\mathbf{g}$ denotes the acceleration of gravity. Combining $\bnabla \times \mathbf{u} = \mathbf{0}$ with Eq.~\eqref{eq:incomp} yields that the scalar velocity potential $\phi$,  defined as $\mathbf{u} = \bnabla \phi$ satisfies the Laplace equation: 
				\begin{eqnarray}
				\Delta \phi = 0\, .\label{eq:laplace}
			\end{eqnarray}
			Equations~\eqref{euler_equation} and~\eqref{eq:laplace} need to be complemented by the linearized boundary conditions at $z = \zeta$: 
			\begin{subeqnarray}\label{eq:bc}
				\partial_t \zeta &=& \partial_z\phi|_{z = \zeta}\, , \slabel{kinematic} \\
				P|_{z = \zeta} &=& P_{\mathrm{atm}} + P_\text{ext} - \gamma \Delta_{x,y} \zeta\, , \slabel{dynamical}
			\end{subeqnarray}
			respectively called kinematic and dynamical boundary conditions, and where $\gamma$ is the surface tension. Combining Eqs~\eqref{translation} to \eqref{eq:bc} with some Fourier analysis yields a second-order ordinary linear differential equation in time for the Fourier Transform of the surface elevation $\hat\zeta(\mathbf{k}, t)$ (see e.g.~\cite{raphael1996capillary,closa2010capillary}):
	\begin{equation}\label{zeta}
				\partial_{t}^2 \hat\zeta(\mathbf{k}, t) + \omega(k)^2 \hat\zeta(\mathbf{k}, t) = -\frac{1}{\rho}k\hat P_{\text{e}}(\mathbf{k})\text{e}^{- \text{i} \mathbf{k} \bcdot \mathbf{r_0}(t)},
			\end{equation}
			with $\omega(k)^2 = gk + \gamma k^3/\rho$.
			Choosing as initial condition that the disturbance is turned on at $t = 0$ (which implies that $\zeta = 0$ beforehand):
			\begin{subeqnarray}
				\hat \zeta(\mathbf{k}, t = 0) &=& 0 \slabel{c_1}\, , \\
				\partial_t \hat \zeta(\mathbf{k}, t = 0) &=& 0 \slabel{c_2}\, ,
			\end{subeqnarray}	
			one obtains a general solution of Eq.~\eqref{zeta} which holds for all trajectories $\mathbf{r_0}(t)$ (see
{Appendix~\ref{solving_differential_equation}} for the details):
			\begin{equation}
			\hat \zeta(\mathbf{k}, t) = - \int_0^t \sin(\omega(k)(t - \tau)) \frac{k \hat P_\text{e}(\mathbf{k}) \text{e}^{-\text{i} \mathbf{k} \bcdot \mathbf{r_0}(\tau)}}{\rho \omega(k)}  \text{d} \tau\, . \label{general_equation}
			\end{equation}

 
In the case of linear motion
 considered in Closa \textit{et al.}~\cite{closa2010capillary} in which the disturbance undergoes uniform linear straight motion  $\mathbf{r_0}(t) = r_0(t) \mathbf{u_x}$ with $r_0(t)=v t$ starting at $t=0$, one can show that the wake converges to the well-known Havelock steady wave pattern. Using Eq.~\eqref{general_equation} one obtains the surface elevation $\hat \zeta_\text{d}(\mathbf{k}, t)$ in the frame of reference attached to the moving disturbance (see {Appendix~\ref{computation_zeta_hat}} for the details):
			\begin{align}
				\hat \zeta_\text{d}(\mathbf{k}, t) &= - \frac{k \hat P_\text{e}(\mathbf{k})}{\rho \omega(k)} \frac{1}{\omega(k)^2 - (v k_x)^2} 
				\left[\omega(k) - \text{e}^{\text{i} v k_x t} \cos(\omega(k) t) \omega(k) + \text{i} \text{e}^{\text{i} v k_x t} \sin(\omega(k) t) v k_x\right]. \label{zeta_hat_expression}
			\end{align}					
			Breaking up the cosine and sine functions into terms of the form $\text{e}^{\pm \text{i} \omega t}$, one can write $\hat \zeta_\text{d}(\mathbf{k}, t)$ as the sum of five terms $\hat \zeta_{\text{d}, j}(\mathbf{k}, t)$ for $j \in [1, 5]$, the first one ($j=1$) being the constant one while the four others are oscillating functions of time. Noticing incidentally that 
			$\hat \zeta_{\text{d}, 1}(\mathbf{k}, t) = \hat \zeta_{\text{st}}(\mathbf{k})$, the well-known steady wave pattern, see e.g.~\cite{raphael1996capillary,closa2010capillary}:
						\begin{align}
				\hat \zeta_\text{st}(\mathbf{k}) &= - \frac{1}{\rho}\frac{k \hat P_\text{e}(\mathbf{k})}{\omega(k)^2 - (v k_x)^2}\, 
		. \label{}
			\end{align}	
		To prove that the unsteady wake converges to its steady counterpart, one then needs to prove that the four other contributions vanish as $t \to +\infty$. In {Appendix~\ref{convergence}}, we show that for all $j\geqslant 2$, 
				$\zeta_{\text{d}, j}(\mathbf{r}, t) \to 0$, and as a result $
				{\zeta_\text{d}(\mathbf{r}, t) \to \zeta_{\text{st}}(\mathbf{r})}$.	
	\medskip
	
	Such results extend to any linear motion for which a constant final velocity is reached in finite time, see {Appendix~\ref{convergence}}. A more realistic  trajectory would indeed consist of an acceleration phase during which the velocity smoothly increases until it reaches a cruising plateau.

	\section{Unsteady Wave Drag}
			
		As mentioned above, the energy transferred to the waves translates into a drag force exerted on the disturbance and opposing its motion. In the case of steady motion and  according to Havelock's reasoning in \cite{havelock1932theory}, the wave drag $\mathbf R_\mathrm{w}$ is given by the total resolved pressure in the direction of the motion $
				P_\text{e}\text{d}\mathbf{S}\bcdot\mathbf{m}$ where $\mathbf m$ is the unit vector collinear to the velocity, and $\text{d}\mathbf{S}$ the surface element vector orthogonal to the air-water interface. Using that $\text{d}\mathbf{S}\bcdot\mathbf{m} = (\mathbf m \bcdot \bnabla) \zeta  \mathbf{m} \text{d}S$, one obtains Havelock's formula:
			\begin{equation}
				\mathbf{R_{\mathrm{w}}}=-\iint P_\text{e}(\mathbf m \bcdot \bnabla)  \zeta  \mathbf{m}\text{d}S\, .
				\label{HavelockF}
			\end{equation}
Such a result can also be obtained from a simple energy balance. The power transferred to the waves by the moving disturbance writes~\cite{wehausen1960surface}:
		\begin{equation}
				\mathcal{P}=\iint P_\text{e}\nabla_n\phi  \text{d}S \ ,
			\label{Laitone}
			\end{equation}
					with $\nabla_n$ denoting the derivative along the vector normal to the surface.
Noting that $\nabla_n\phi=\mathbf v \bcdot\bnabla\zeta$ where $\mathbf v = \partial_t{ \mathbf{r_0}}$, and matching	$\mathcal{P}$ 	to the power of a force acting against the motion, $ \mathcal{P} = -	\mathbf R_{\mathrm{w}}\bcdot \mathbf v $, yields  Eq.~\eqref{HavelockF}.\medskip

Extending such reasoning to the unsteady setting in which the surface elevation is time-dependent yields a nonzero vertical component.  Using Havelock's approach, the horizontal component is given as before by:
			\begin{equation}
				\mathcal{P}_w(t)=-\iint P_\text{e}(\mathbf v \bcdot \bnabla) \zeta \text{d}S \ .
			\end{equation}
			The power of the vertical component can, on the other hand, be computed by noting that the infinitesimal vertical work of the pressure field on the surface between $t$ and $t+\text{d}t$ is given by $\delta W_v=-P_\text{e} \text dz$ with $\text dz =\partial_t\zeta(x,y,t)\,\text{d}t$ the infinitesimal vertical displacement of the surface.  The vertical power thus reads:
			\begin{equation}
				\mathcal{P}_v(t)=-\iint P_\text{e}  \partial_t\zeta  \text{d}S\, .
			\end{equation}
			This term naturally vanishes in the steady case 
 ($\partial_t\zeta=0$). Multiplying the kinematic boundary condition in the frame of reference of the moving disturbance $\partial_t \zeta +(\mathbf{v}\cdot\bnabla)\zeta=\partial_z\phi $ by $P_e$ and integrating over the surface, one sees that the overall power transferred to the waves matches the power of the vertical and horizontal forces acting on the moving disturbance: $				\mathcal{P}=-(\mathcal{P}_w+\mathcal{P}_v) 
$,	with $\mathcal{P}(t)$ given by Eq.~\eqref{Laitone}. \medskip

In summary, for an arbitrarily moving disturbance the power transferred to the waves translates into a force with a vertical and a horizontal components. The horizontal one is what we call wave drag (at standstill, the moving disturbance does not experience any drag, only vertical oscillations~\cite{closa2010capillary}). Combining Eq.~\eqref{HavelockF}  with Eq.~\eqref{general_equation} yields:
			\begin{eqnarray}
	\mathbf R_{\mathrm{w}} &=&  \frac{\mathbf m}{4\pi^2} \iint {\text d^2 \mathbf k} \; \frac{\text{i}k |\hat P_\text{e}(\mathbf k)|^2(\mathbf k \bcdot \mathbf m )}{ \rho \omega(k)}  \int_0^t \text d \tau \sin(\omega(k) (t - \tau)) \text e^{\text{i} \mathbf k \bcdot (\mathbf r_0(t) - \mathbf r_0(\tau))}
\,  , \label{draggen}
\end{eqnarray}
where we recall that $\mathbf m$ is the unit vector collinear to the velocity.
Eq.~\eqref{draggen} is the central result of this paper. It gives the instantaneous wave drag for any trajectory $\mathbf{r_0}(t)$ with nonzero velocity. In the particular case of linear motion, and axi-symmetric pressure disturbance, $\hat P_\text{e}(\mathbf k)=\hat P_\text{e}( k)$, Eq.~\eqref{draggen} simplifies to:
\begin{eqnarray}
\mathbf R_{\mathrm{w}} &=& -\frac{\mathbf m}{2\pi}\int_0^\infty\text{d}k\frac{k^3\vert\hat P_\text{e}( k)\vert^2}{\rho \omega(k)}\int_0^t\text{d}\tau\sin(\omega(k)(t-\tau))J_1( k ( r_0(t)- r_0(\tau)))\, , \label{draglin}
\end{eqnarray}
where $J_1$ denotes the Bessel function of the first kind and of order 1. 
Note that in the case of a linear sudden object motion described above one recovers the results of Closa \textit{et al.}~\cite{closa2010capillary}.
 Also note that in the more general case of linear motion with constant final velocity reached in finite time, having shown that the wake pattern in the disturbance's frame of reference converges to a constant, the same is true concerning wave drag.
			\section{How slow is slow enough?}

			\begin{figure}
				\centering
				\includegraphics[width=\textwidth]{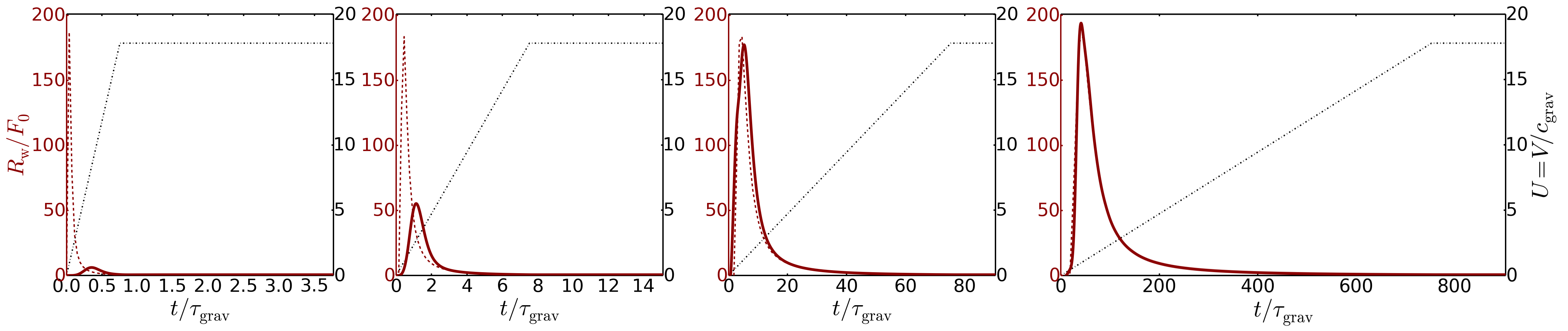}
				\caption{Pure gravity instantaneous wave drag (dark red curves and left vertical axis) experienced by a moving disturbance for four different values of the acceleration -- from left to right:  $a\approx 230$, $23$, $2.3$ and $0.23$\,m$\cdot$s$^{-2}$ -- as function of time rescaled by $\tau_\text{grav}=\sqrt{{b}/{g}}$. The dashed dark red line shows the wave drag computed from Havelock's steady wave drag formula. The dash-dotted black line signifies the velocity profile rescaled by $c_\text{grav}=\sqrt{g b}$ (right vertical axis). }
				\label{graphs_g}
			\end{figure}

			In this section we raise the question of the accuracy of using Havelock's steady formula to compute the wave drag in an unsteady situation. In other words,	how small does the acceleration need to be for the evolution to be reasonably considered as quasi-static wave drag-wise. 
			To answer this question in a stylized setting, we compute $R_{\mathrm{w}}(t)$ for various different velocity profiles. More precisely we choose a ramp velocity profile with constant acceleration  of the form $\mathbf{r_0}(t) =\frac12 a t^2 \mathbf{u_x}$.\medskip
			
			Figure~\ref{graphs_g} shows the result in the pure gravity limit, $\omega(k)^2=gk$, for four values of the acceleration $a$, computed with a  Lorentzian pressure field of the form $ \hat{P}_\text{e}(k) = F_0 \mathrm{e}^{- b |k|}$,
			where $b$ is the typical size of the disturbance. One can see that, as expected, the quasi-steady limit is achieved for $a\ll g$.
			\medskip
			
\begin{figure}
				\centering
				\includegraphics[width=\textwidth]{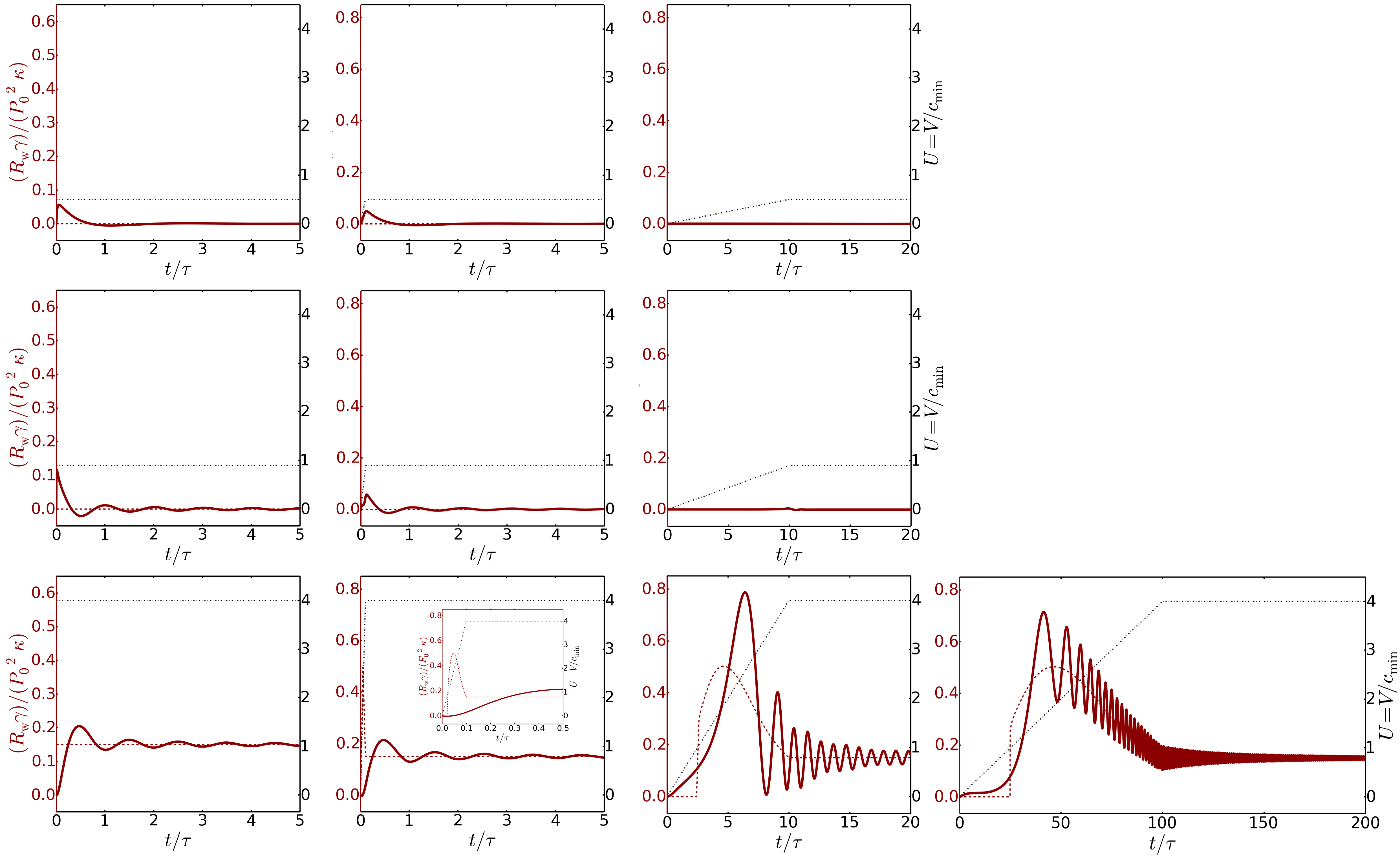}
				\caption{
				Capillary-gravity instantaneous wave drag (dark red curves and left vertical axis) experienced by a moving disturbance for three values of the final velocity -- from top to bottom: $v_\infty/c_\text{min}= 0.5$, $0.9$  and $4$ where $c_\text{min} = \left(4 g \rho / \gamma \right)^{1/4}\approx 0.23$~m$\cdot$s$^{-1}$ -- and four different durations of the velocity ramp. -- from left to right:  $t_\mathrm{ramp}= 0$ (constant velocity), $\tau/10$ , $10 \tau$ and $100 \tau$ where $\tau$ is the pseudo-period of oscillations. The corresponding accelerations are, from left to right: for $v_\infty/c_\text{min}= 0.5$, $a \approx \infty$, $13$ and $0.13 $~m$\cdot$s$^{-2}$; for $v_\infty/c_\text{min}= 0.9$, $a \approx \infty$, $2.5$ and $0.025$~m$\cdot$s$^{-2}$; for $v_\infty/c_\text{min}= 4$, $a \approx \infty$, $2300$, $23$ and $2.3$~m$\cdot$s$^{-2}$. The dashed dark red line shows the wave drag computed from Havelock's steady wave drag formula. The dash-dotted black line signifies the velocity profile rescaled $c_\text{min}$ (right vertical axis).  The two seemingly missing panels have been left out as they require unnecessary computing power, in addition to being almost perfectly flat lines at $R_\mathrm{w}=0$.
	}
				\label{graphs_cg}
				
			\end{figure}

			Figure~\ref{graphs_cg} shows the results for capillary-gravity waves for three values of the final velocity $v_\infty$ and four values of the velocity ramp's duration $t_\mathrm{ramp}$. We choose $b={\kappa^{-1}}/{10}$, with $\kappa^{-1} = \sqrt{{\gamma}/({\rho g})}$ the capillary length, to ensure that capillary effects are non-negligible~\cite{benzaquen2011wave}. 
		While we consistently find again that the instantaneous wave drag coincides better and better with its steady counterpart as the acceleration is decreased, significant oscillations remain, at odds with the pure gravity case. 
	The three rows describe three physically different regimes of final velocities, see~\cite{closa2010capillary}. For $v_\infty < c_\mathrm{min}$, the final wave drag is zero, whereas it is non-zero whenever $v_\infty \geqslant c_\mathrm{min}$. Further, while for 
			 $v_\infty < c_\mathrm{crit}\approx 0.77c_\mathrm{min}$ the oscillations decay exponentially, for $v_\infty > c_\mathrm{crit}$ they decay as $1/t$. \medskip

	Also note that even for the weakest acceleration (bottom right panel in Fig.~\ref{graphs_cg}), the discontinuity of wave drag in the capillary-gravity case expected to occur at $U/c_{\min} = 0.23$, as computed by Rapha\"el \& de Gennes~\cite{raphael1996capillary}, is phased out in the unsteady case. This could be a solution to the apparent contradiction with the experimental results of Burghelea \& Steinberg~\cite{burghelea2002wave,burghelea2001onset}, who claimed that such discontinuity did not exist. See also  \cite{richard1999capillary} and \cite{benzaquen2011wave} for alternative explanations.	

	\section{Conclusion}
	
	In this paper we have derived a general formula to compute the instantaneous wave drag exerted on an arbitrarily moving disturbance. In particular, we have assessed in a specific situation to which extent one can rightfully use Havelock's steady wave drag formula. Such an approach can be used to legitimize or discredit a number of studies which used steady wave drag formulas in unsteady situations, see e.g.~\cite{le2011wave,benham2020hysteretic}. 
	\medskip

 Of particular interest is the experimental analysis of Le Merrer \textit{et al.}~\cite{le2011wave} in which the capillary-gravity wave drag was inferred from the 
free deceleration of a liquid nitrogen droplet launched over the water surface~\cite{le2011wave}. The discrepancies observed by the authors between the experiments and the theoretical steady wave drag could perhaps be attributed to the fact that their experiments did not fall in the quasi-static regime in which steady Havelock is accurate. To solve this free  deceleration problem  one needs to compute jointly the wave drag (Eq.~\eqref{draglin}) and the resulting dynamics of the moving disturbance (the nitrogen droplet) from Newton's law: $m \ddot r_0=-\beta\dot r_0 -R_{\mathrm{w}}(\lbrace r_0(t)\rbrace)$, 
		with $m$ the mass of the liquid nitrogen droplet and $\beta$ a friction coefficient.
Further, computing the average wave drag for an oscillating velocity profile of the form $v(t) = v_0(1 + \epsilon \sin{\Omega t})$ would be highly relevant to address optimal strategies in rowing sports, see~\cite{boucher2017row, labbe2019physics}. Such studies are left for future work.\medskip

	We thank  C. Clanet, A. Darmon, U. Mizrahi and F. Vandenbrouck for fruitful discussions. We also thank Z. Zeravcic for her help with the numerical computations, and R. Carmigniani for his thorough proofreading.
				

    \bibliography{biblio}

\begin{thebibliography}{27}%
\makeatletter
\providecommand \@ifxundefined [1]{%
 \@ifx{#1\undefined}
}%
\providecommand \@ifnum [1]{%
 \ifnum #1\expandafter \@firstoftwo
 \else \expandafter \@secondoftwo
 \fi
}%
\providecommand \@ifx [1]{%
 \ifx #1\expandafter \@firstoftwo
 \else \expandafter \@secondoftwo
 \fi
}%
\providecommand \natexlab [1]{#1}%
\providecommand \enquote  [1]{``#1''}%
\providecommand \bibnamefont  [1]{#1}%
\providecommand \bibfnamefont [1]{#1}%
\providecommand \citenamefont [1]{#1}%
\providecommand \href@noop [0]{\@secondoftwo}%
\providecommand \href [0]{\begingroup \@sanitize@url \@href}%
\providecommand \@href[1]{\@@startlink{#1}\@@href}%
\providecommand \@@href[1]{\endgroup#1\@@endlink}%
\providecommand \@sanitize@url [0]{\catcode `\\12\catcode `\$12\catcode
  `\&12\catcode `\#12\catcode `\^12\catcode `\_12\catcode `\%12\relax}%
\providecommand \@@startlink[1]{}%
\providecommand \@@endlink[0]{}%
\providecommand \url  [0]{\begingroup\@sanitize@url \@url }%
\providecommand \@url [1]{\endgroup\@href {#1}{\urlprefix }}%
\providecommand \urlprefix  [0]{URL }%
\providecommand \Eprint [0]{\href }%
\providecommand \doibase [0]{http://dx.doi.org/}%
\providecommand \selectlanguage [0]{\@gobble}%
\providecommand \bibinfo  [0]{\@secondoftwo}%
\providecommand \bibfield  [0]{\@secondoftwo}%
\providecommand \translation [1]{[#1]}%
\providecommand \BibitemOpen [0]{}%
\providecommand \bibitemStop [0]{}%
\providecommand \bibitemNoStop [0]{.\EOS\space}%
\providecommand \EOS [0]{\spacefactor3000\relax}%
\providecommand \BibitemShut  [1]{\csname bibitem#1\endcsname}%
\let\auto@bib@innerbib\@empty
\bibitem [{\citenamefont {Darrigol}(2005)}]{darrigol2005worlds}%
  \BibitemOpen
  \bibfield  {author} {\bibinfo {author} {\bibfnamefont {Olivier}\ \bibnamefont
  {Darrigol}},\ }\href@noop {} {\emph {\bibinfo {title} {Worlds of flow: A
  history of hydrodynamics from the Bernoullis to Prandtl}}}\ (\bibinfo
  {publisher} {Oxford University Press},\ \bibinfo {year} {2005})\BibitemShut
  {NoStop}%
\bibitem [{\citenamefont {Lighthill}(1978)}]{lighthill1978waves}%
  \BibitemOpen
  \bibfield  {author} {\bibinfo {author} {\bibfnamefont {J}~\bibnamefont
  {Lighthill}},\ }\href@noop {} {\emph {\bibinfo {title} {Waves in Fluids}}}\
  (\bibinfo {year} {1978})\ p.\ \bibinfo {pages} {120}\BibitemShut {NoStop}%
\bibitem [{\citenamefont {Kelvin}(1887)}]{kelvin1887ship}%
  \BibitemOpen
  \bibfield  {author} {\bibinfo {author} {\bibfnamefont {Lord}\ \bibnamefont
  {Kelvin}},\ }\bibfield  {title} {\enquote {\bibinfo {title} {On ship
  waves},}\ }\href@noop {} {\bibfield  {journal} {\bibinfo  {journal} {Proc.
  Inst. Mech. Eng}\ }\textbf {\bibinfo {volume} {3}},\ \bibinfo {pages}
  {409--434} (\bibinfo {year} {1887})}\BibitemShut {NoStop}%
\bibitem [{\citenamefont {Rabaud}\ and\ \citenamefont
  {Moisy}(2013)}]{rabaud2013ship}%
  \BibitemOpen
  \bibfield  {author} {\bibinfo {author} {\bibfnamefont {Marc}\ \bibnamefont
  {Rabaud}}\ and\ \bibinfo {author} {\bibfnamefont {Fr{\'e}d{\'e}ric}\
  \bibnamefont {Moisy}},\ }\bibfield  {title} {\enquote {\bibinfo {title} {Ship
  wakes: Kelvin or {M}ach angle?}}\ }\href@noop {} {\bibfield  {journal}
  {\bibinfo  {journal} {Physical Review Letters}\ }\textbf {\bibinfo {volume}
  {110}},\ \bibinfo {pages} {214503} (\bibinfo {year} {2013})}\BibitemShut
  {NoStop}%
\bibitem [{\citenamefont {Darmon}\ \emph {et~al.}(2014)\citenamefont {Darmon},
  \citenamefont {Benzaquen},\ and\ \citenamefont
  {Rapha{\"e}l}}]{darmon2014kelvin}%
  \BibitemOpen
  \bibfield  {author} {\bibinfo {author} {\bibfnamefont {Alexandre}\
  \bibnamefont {Darmon}}, \bibinfo {author} {\bibfnamefont {Michael}\
  \bibnamefont {Benzaquen}}, \ and\ \bibinfo {author} {\bibfnamefont
  {{\'E}lie}\ \bibnamefont {Rapha{\"e}l}},\ }\bibfield  {title} {\enquote
  {\bibinfo {title} {Kelvin wake pattern at large {F}roude numbers},}\
  }\href@noop {} {\bibfield  {journal} {\bibinfo  {journal} {Journal of Fluid
  Mechanics}\ }\textbf {\bibinfo {volume} {738}} (\bibinfo {year}
  {2014})}\BibitemShut {NoStop}%
\bibitem [{\citenamefont {Wehausen}\ and\ \citenamefont
  {Laitone}(1960)}]{wehausen1960surface}%
  \BibitemOpen
  \bibfield  {author} {\bibinfo {author} {\bibfnamefont {John~V}\ \bibnamefont
  {Wehausen}}\ and\ \bibinfo {author} {\bibfnamefont {Edmund~V}\ \bibnamefont
  {Laitone}},\ }\bibfield  {title} {\enquote {\bibinfo {title} {Surface
  waves},}\ }in\ \href@noop {} {\emph {\bibinfo {booktitle} {Fluid
  Dynamics/Str{\"o}mungsmechanik}}}\ (\bibinfo  {publisher} {Springer},\
  \bibinfo {year} {1960})\ pp.\ \bibinfo {pages} {446--778}\BibitemShut
  {NoStop}%
\bibitem [{\citenamefont {Havelock}(1932)}]{havelock1932theory}%
  \BibitemOpen
  \bibfield  {author} {\bibinfo {author} {\bibfnamefont {TH}~\bibnamefont
  {Havelock}},\ }\bibfield  {title} {\enquote {\bibinfo {title} {The theory of
  wave resistance},}\ }\href@noop {} {\bibfield  {journal} {\bibinfo  {journal}
  {Proceedings of the Royal Society of London. Series A, Containing Papers of a
  Mathematical and Physical Character}\ }\textbf {\bibinfo {volume} {138}},\
  \bibinfo {pages} {339--348} (\bibinfo {year} {1932})}\BibitemShut {NoStop}%
\bibitem [{\citenamefont {Michell}(1898)}]{michell1898xi}%
  \BibitemOpen
  \bibfield  {author} {\bibinfo {author} {\bibfnamefont {John~Henry}\
  \bibnamefont {Michell}},\ }\bibfield  {title} {\enquote {\bibinfo {title}
  {Xi. the wave-resistance of a ship},}\ }\href@noop {} {\bibfield  {journal}
  {\bibinfo  {journal} {The London, Edinburgh, and Dublin Philosophical
  Magazine and Journal of Science}\ }\textbf {\bibinfo {volume} {45}},\
  \bibinfo {pages} {106--123} (\bibinfo {year} {1898})}\BibitemShut {NoStop}%
\bibitem [{\citenamefont {Parnell}\ and\ \citenamefont
  {Kofoed-Hansen}(2001)}]{parnell2001wakes}%
  \BibitemOpen
  \bibfield  {author} {\bibinfo {author} {\bibfnamefont {Kevin~E}\ \bibnamefont
  {Parnell}}\ and\ \bibinfo {author} {\bibfnamefont {Henrik}\ \bibnamefont
  {Kofoed-Hansen}},\ }\bibfield  {title} {\enquote {\bibinfo {title} {Wakes
  from large high-speed ferries in confined coastal waters: Management
  approaches with examples from {N}ew {Z}ealand and {D}enmark},}\ }\href@noop
  {} {\bibfield  {journal} {\bibinfo  {journal} {Coastal Management}\ }\textbf
  {\bibinfo {volume} {29}},\ \bibinfo {pages} {217--237} (\bibinfo {year}
  {2001})}\BibitemShut {NoStop}%
\bibitem [{\citenamefont {Benzaquen}\ \emph {et~al.}(2014)\citenamefont
  {Benzaquen}, \citenamefont {Darmon},\ and\ \citenamefont
  {Rapha{\"e}l}}]{benzaquen2014wake}%
  \BibitemOpen
  \bibfield  {author} {\bibinfo {author} {\bibfnamefont {Michael}\ \bibnamefont
  {Benzaquen}}, \bibinfo {author} {\bibfnamefont {Alexandre}\ \bibnamefont
  {Darmon}}, \ and\ \bibinfo {author} {\bibfnamefont {{\'E}lie}\ \bibnamefont
  {Rapha{\"e}l}},\ }\bibfield  {title} {\enquote {\bibinfo {title} {Wake
  pattern and wave resistance for anisotropic moving disturbances},}\
  }\href@noop {} {\bibfield  {journal} {\bibinfo  {journal} {Physics of
  Fluids}\ }\textbf {\bibinfo {volume} {26}},\ \bibinfo {pages} {092106}
  (\bibinfo {year} {2014})}\BibitemShut {NoStop}%
\bibitem [{\citenamefont {Boucher}\ \emph {et~al.}(2018)\citenamefont
  {Boucher}, \citenamefont {Labb{\'e}}, \citenamefont {Clanet},\ and\
  \citenamefont {Benzaquen}}]{boucher2018thin}%
  \BibitemOpen
  \bibfield  {author} {\bibinfo {author} {\bibfnamefont {Jean-Philippe}\
  \bibnamefont {Boucher}}, \bibinfo {author} {\bibfnamefont {Romain}\
  \bibnamefont {Labb{\'e}}}, \bibinfo {author} {\bibfnamefont {Christophe}\
  \bibnamefont {Clanet}}, \ and\ \bibinfo {author} {\bibfnamefont {Michael}\
  \bibnamefont {Benzaquen}},\ }\bibfield  {title} {\enquote {\bibinfo {title}
  {Thin or bulky: Optimal aspect ratios for ship hulls},}\ }\href@noop {}
  {\bibfield  {journal} {\bibinfo  {journal} {Physical Review Fluids}\ }\textbf
  {\bibinfo {volume} {3}},\ \bibinfo {pages} {074802} (\bibinfo {year}
  {2018})}\BibitemShut {NoStop}%
\bibitem [{\citenamefont {Benham}\ \emph {et~al.}(2019)\citenamefont {Benham},
  \citenamefont {Boucher}, \citenamefont {Labb{\'e}}, \citenamefont
  {Benzaquen},\ and\ \citenamefont {Clanet}}]{benham2019wave}%
  \BibitemOpen
  \bibfield  {author} {\bibinfo {author} {\bibfnamefont {GP}~\bibnamefont
  {Benham}}, \bibinfo {author} {\bibfnamefont {Jean-Philippe}\ \bibnamefont
  {Boucher}}, \bibinfo {author} {\bibfnamefont {Romain}\ \bibnamefont
  {Labb{\'e}}}, \bibinfo {author} {\bibfnamefont {Michael}\ \bibnamefont
  {Benzaquen}}, \ and\ \bibinfo {author} {\bibfnamefont {Christophe}\
  \bibnamefont {Clanet}},\ }\bibfield  {title} {\enquote {\bibinfo {title}
  {Wave drag on asymmetric bodies},}\ }\href@noop {} {\bibfield  {journal}
  {\bibinfo  {journal} {Journal of Fluid Mechanics}\ }\textbf {\bibinfo
  {volume} {878}},\ \bibinfo {pages} {147--168} (\bibinfo {year}
  {2019})}\BibitemShut {NoStop}%
\bibitem [{\citenamefont {Benham}\ \emph {et~al.}(2020)\citenamefont {Benham},
  \citenamefont {Bendimerad}, \citenamefont {Benzaquen},\ and\ \citenamefont
  {Clanet}}]{benham2020hysteretic}%
  \BibitemOpen
  \bibfield  {author} {\bibinfo {author} {\bibfnamefont {GP}~\bibnamefont
  {Benham}}, \bibinfo {author} {\bibfnamefont {R}~\bibnamefont {Bendimerad}},
  \bibinfo {author} {\bibfnamefont {M}~\bibnamefont {Benzaquen}}, \ and\
  \bibinfo {author} {\bibfnamefont {C}~\bibnamefont {Clanet}},\ }\bibfield
  {title} {\enquote {\bibinfo {title} {Hysteretic wave drag in shallow
  water},}\ }\href@noop {} {\bibfield  {journal} {\bibinfo  {journal} {arXiv
  preprint arXiv:2003.00502}\ } (\bibinfo {year} {2020})}\BibitemShut {NoStop}%
\bibitem [{\citenamefont {Rapha{\"e}l}\ and\ \citenamefont
  {De~Gennes}(1996)}]{raphael1996capillary}%
  \BibitemOpen
  \bibfield  {author} {\bibinfo {author} {\bibfnamefont {E.}~\bibnamefont
  {Rapha{\"e}l}}\ and\ \bibinfo {author} {\bibfnamefont {P.-G.}\ \bibnamefont
  {De~Gennes}},\ }\bibfield  {title} {\enquote {\bibinfo {title} {Capillary
  gravity waves caused by a moving disturbance: wave resistance},}\ }\href@noop
  {} {\bibfield  {journal} {\bibinfo  {journal} {Physical Review E}\ }\textbf
  {\bibinfo {volume} {53}},\ \bibinfo {pages} {3448} (\bibinfo {year}
  {1996})}\BibitemShut {NoStop}%
\bibitem [{\citenamefont {Voise}\ and\ \citenamefont
  {Casas}(2010)}]{voise2010management}%
  \BibitemOpen
  \bibfield  {author} {\bibinfo {author} {\bibfnamefont {Jonathan}\
  \bibnamefont {Voise}}\ and\ \bibinfo {author} {\bibfnamefont
  {J{\'e}r{\^o}me}\ \bibnamefont {Casas}},\ }\bibfield  {title} {\enquote
  {\bibinfo {title} {The management of fluid and wave resistances by whirligig
  beetles},}\ }\href@noop {} {\bibfield  {journal} {\bibinfo  {journal}
  {Journal of The Royal Society Interface}\ }\textbf {\bibinfo {volume} {7}},\
  \bibinfo {pages} {343--352} (\bibinfo {year} {2010})}\BibitemShut {NoStop}%
\bibitem [{\citenamefont {Hu}\ \emph {et~al.}(2003)\citenamefont {Hu},
  \citenamefont {Chan},\ and\ \citenamefont {Bush}}]{hu2003hydrodynamics}%
  \BibitemOpen
  \bibfield  {author} {\bibinfo {author} {\bibfnamefont {David~L}\ \bibnamefont
  {Hu}}, \bibinfo {author} {\bibfnamefont {Brian}\ \bibnamefont {Chan}}, \ and\
  \bibinfo {author} {\bibfnamefont {John~WM}\ \bibnamefont {Bush}},\ }\bibfield
   {title} {\enquote {\bibinfo {title} {The hydrodynamics of water strider
  locomotion},}\ }\href@noop {} {\bibfield  {journal} {\bibinfo  {journal}
  {Nature}\ }\textbf {\bibinfo {volume} {424}},\ \bibinfo {pages} {663--666}
  (\bibinfo {year} {2003})}\BibitemShut {NoStop}%
\bibitem [{\citenamefont {B{\"u}hler}(2007)}]{buhler2007impulsive}%
  \BibitemOpen
  \bibfield  {author} {\bibinfo {author} {\bibfnamefont {Oliver}\ \bibnamefont
  {B{\"u}hler}},\ }\bibfield  {title} {\enquote {\bibinfo {title} {Impulsive
  fluid forcing and water strider locomotion},}\ }\href@noop {} {\bibfield
  {journal} {\bibinfo  {journal} {Journal of Fluid Mechanics}\ }\textbf
  {\bibinfo {volume} {573}},\ \bibinfo {pages} {211--236} (\bibinfo {year}
  {2007})}\BibitemShut {NoStop}%
\bibitem [{\citenamefont {Steinmann}\ \emph {et~al.}(2018)\citenamefont
  {Steinmann}, \citenamefont {Arutkin}, \citenamefont {Cochard}, \citenamefont
  {Rapha{\"e}l}, \citenamefont {Casas},\ and\ \citenamefont
  {Benzaquen}}]{steinmann2018unsteady}%
  \BibitemOpen
  \bibfield  {author} {\bibinfo {author} {\bibfnamefont {Thomas}\ \bibnamefont
  {Steinmann}}, \bibinfo {author} {\bibfnamefont {Maxence}\ \bibnamefont
  {Arutkin}}, \bibinfo {author} {\bibfnamefont {Pr{\'e}cillia}\ \bibnamefont
  {Cochard}}, \bibinfo {author} {\bibfnamefont {{\'E}lie}\ \bibnamefont
  {Rapha{\"e}l}}, \bibinfo {author} {\bibfnamefont {J{\'e}r{\^o}me}\
  \bibnamefont {Casas}}, \ and\ \bibinfo {author} {\bibfnamefont {Michael}\
  \bibnamefont {Benzaquen}},\ }\bibfield  {title} {\enquote {\bibinfo {title}
  {Unsteady wave pattern generation by water striders},}\ }\href@noop {}
  {\bibfield  {journal} {\bibinfo  {journal} {Journal of Fluid Mechanics}\
  }\textbf {\bibinfo {volume} {848}},\ \bibinfo {pages} {370--387} (\bibinfo
  {year} {2018})}\BibitemShut {NoStop}%
\bibitem [{\citenamefont {Le~Merrer}\ \emph {et~al.}(2011)\citenamefont
  {Le~Merrer}, \citenamefont {Clanet}, \citenamefont {Qu{\'e}r{\'e}},
  \citenamefont {Rapha{\"e}l},\ and\ \citenamefont {Chevy}}]{le2011wave}%
  \BibitemOpen
  \bibfield  {author} {\bibinfo {author} {\bibfnamefont {Marie}\ \bibnamefont
  {Le~Merrer}}, \bibinfo {author} {\bibfnamefont {Christophe}\ \bibnamefont
  {Clanet}}, \bibinfo {author} {\bibfnamefont {David}\ \bibnamefont
  {Qu{\'e}r{\'e}}}, \bibinfo {author} {\bibfnamefont {{\'E}lie}\ \bibnamefont
  {Rapha{\"e}l}}, \ and\ \bibinfo {author} {\bibfnamefont {Fr{\'e}d{\'e}ric}\
  \bibnamefont {Chevy}},\ }\bibfield  {title} {\enquote {\bibinfo {title} {Wave
  drag on floating bodies},}\ }\href@noop {} {\bibfield  {journal} {\bibinfo
  {journal} {Proceedings of the National Academy of Sciences}\ }\textbf
  {\bibinfo {volume} {108}},\ \bibinfo {pages} {15064--15068} (\bibinfo {year}
  {2011})}\BibitemShut {NoStop}%
\bibitem [{\citenamefont {Dutykh}\ and\ \citenamefont
  {Dias}(2007)}]{dutykh2007water}%
  \BibitemOpen
  \bibfield  {author} {\bibinfo {author} {\bibfnamefont {Denys}\ \bibnamefont
  {Dutykh}}\ and\ \bibinfo {author} {\bibfnamefont {Fr{\'e}d{\'e}ric}\
  \bibnamefont {Dias}},\ }\bibfield  {title} {\enquote {\bibinfo {title} {Water
  waves generated by a moving bottom},}\ }in\ \href@noop {} {\emph {\bibinfo
  {booktitle} {Tsunami and Nonlinear waves}}}\ (\bibinfo  {publisher}
  {Springer},\ \bibinfo {year} {2007})\ pp.\ \bibinfo {pages}
  {65--95}\BibitemShut {NoStop}%
\bibitem [{\citenamefont {Closa}\ \emph {et~al.}(2010)\citenamefont {Closa},
  \citenamefont {Chepelianskii},\ and\ \citenamefont
  {Rapha{\"e}l}}]{closa2010capillary}%
  \BibitemOpen
  \bibfield  {author} {\bibinfo {author} {\bibfnamefont {Fabien}\ \bibnamefont
  {Closa}}, \bibinfo {author} {\bibfnamefont {AD}~\bibnamefont
  {Chepelianskii}}, \ and\ \bibinfo {author} {\bibfnamefont {{\'E}lie}\
  \bibnamefont {Rapha{\"e}l}},\ }\bibfield  {title} {\enquote {\bibinfo {title}
  {Capillary-gravity waves generated by a sudden object motion},}\ }\href@noop
  {} {\bibfield  {journal} {\bibinfo  {journal} {Physics of Fluids}\ }\textbf
  {\bibinfo {volume} {22}},\ \bibinfo {pages} {052107} (\bibinfo {year}
  {2010})}\BibitemShut {NoStop}%
\bibitem [{\citenamefont {Benzaquen}\ \emph {et~al.}(2011)\citenamefont
  {Benzaquen}, \citenamefont {Chevy},\ and\ \citenamefont
  {Rapha{\"e}l}}]{benzaquen2011wave}%
  \BibitemOpen
  \bibfield  {author} {\bibinfo {author} {\bibfnamefont {Michael}\ \bibnamefont
  {Benzaquen}}, \bibinfo {author} {\bibfnamefont {Fr{\'e}d{\'e}ric}\
  \bibnamefont {Chevy}}, \ and\ \bibinfo {author} {\bibfnamefont {{\'E}lie}\
  \bibnamefont {Rapha{\"e}l}},\ }\bibfield  {title} {\enquote {\bibinfo {title}
  {Wave resistance for capillary gravity waves: Finite-size effects},}\
  }\href@noop {} {\bibfield  {journal} {\bibinfo  {journal} {EPL (Europhysics
  Letters)}\ }\textbf {\bibinfo {volume} {96}},\ \bibinfo {pages} {34003}
  (\bibinfo {year} {2011})}\BibitemShut {NoStop}%
\bibitem [{\citenamefont {Burghelea}\ and\ \citenamefont
  {Steinberg}(2002)}]{burghelea2002wave}%
  \BibitemOpen
  \bibfield  {author} {\bibinfo {author} {\bibfnamefont {Teodor}\ \bibnamefont
  {Burghelea}}\ and\ \bibinfo {author} {\bibfnamefont {Victor}\ \bibnamefont
  {Steinberg}},\ }\bibfield  {title} {\enquote {\bibinfo {title} {Wave drag due
  to generation of capillary-gravity surface waves},}\ }\href@noop {}
  {\bibfield  {journal} {\bibinfo  {journal} {Physical Review E}\ }\textbf
  {\bibinfo {volume} {66}},\ \bibinfo {pages} {051204} (\bibinfo {year}
  {2002})}\BibitemShut {NoStop}%
\bibitem [{\citenamefont {Burghelea}\ and\ \citenamefont
  {Steinberg}(2001)}]{burghelea2001onset}%
  \BibitemOpen
  \bibfield  {author} {\bibinfo {author} {\bibfnamefont {Teodor}\ \bibnamefont
  {Burghelea}}\ and\ \bibinfo {author} {\bibfnamefont {Victor}\ \bibnamefont
  {Steinberg}},\ }\bibfield  {title} {\enquote {\bibinfo {title} {Onset of wave
  drag due to generation of capillary-gravity waves by a moving object as a
  critical phenomenon},}\ }\href@noop {} {\bibfield  {journal} {\bibinfo
  {journal} {Physical review letters}\ }\textbf {\bibinfo {volume} {86}},\
  \bibinfo {pages} {2557} (\bibinfo {year} {2001})}\BibitemShut {NoStop}%
\bibitem [{\citenamefont {Richard}\ and\ \citenamefont
  {Raphael}(1999)}]{richard1999capillary}%
  \BibitemOpen
  \bibfield  {author} {\bibinfo {author} {\bibfnamefont {Denis}\ \bibnamefont
  {Richard}}\ and\ \bibinfo {author} {\bibfnamefont {Elie}\ \bibnamefont
  {Raphael}},\ }\bibfield  {title} {\enquote {\bibinfo {title}
  {Capillary-gravity waves: The effect of viscosity on the wave resistance},}\
  }\href@noop {} {\bibfield  {journal} {\bibinfo  {journal} {EPL (Europhysics
  Letters)}\ }\textbf {\bibinfo {volume} {48}},\ \bibinfo {pages} {49}
  (\bibinfo {year} {1999})}\BibitemShut {NoStop}%
\bibitem [{\citenamefont {Boucher}\ \emph {et~al.}(2017)\citenamefont
  {Boucher}, \citenamefont {Labb{\'e}},\ and\ \citenamefont
  {Clanet}}]{boucher2017row}%
  \BibitemOpen
  \bibfield  {author} {\bibinfo {author} {\bibfnamefont {Jean-Philippe}\
  \bibnamefont {Boucher}}, \bibinfo {author} {\bibfnamefont {Romain}\
  \bibnamefont {Labb{\'e}}}, \ and\ \bibinfo {author} {\bibfnamefont
  {Christophe}\ \bibnamefont {Clanet}},\ }\bibfield  {title} {\enquote
  {\bibinfo {title} {Row bots},}\ }\href@noop {} {\bibfield  {journal}
  {\bibinfo  {journal} {Physics Today}\ }\textbf {\bibinfo {volume} {70}},\
  \bibinfo {pages} {82} (\bibinfo {year} {2017})}\BibitemShut {NoStop}%
\bibitem [{\citenamefont {Labb{\'e}}\ \emph {et~al.}(2019)\citenamefont
  {Labb{\'e}}, \citenamefont {Boucher}, \citenamefont {Clanet},\ and\
  \citenamefont {Benzaquen}}]{labbe2019physics}%
  \BibitemOpen
  \bibfield  {author} {\bibinfo {author} {\bibfnamefont {Romain}\ \bibnamefont
  {Labb{\'e}}}, \bibinfo {author} {\bibfnamefont {Jean-Philippe}\ \bibnamefont
  {Boucher}}, \bibinfo {author} {\bibfnamefont {Christophe}\ \bibnamefont
  {Clanet}}, \ and\ \bibinfo {author} {\bibfnamefont {Michael}\ \bibnamefont
  {Benzaquen}},\ }\bibfield  {title} {\enquote {\bibinfo {title} {Physics of
  rowing oars},}\ }\href@noop {} {\bibfield  {journal} {\bibinfo  {journal}
  {New Journal of Physics}\ }\textbf {\bibinfo {volume} {21}},\ \bibinfo
  {pages} {093050} (\bibinfo {year} {2019})}\BibitemShut {NoStop}%
\end{thebibliography}%

	\appendix
	
	\small
	
	\section{Unsteady wake solution}\label{solving_differential_equation}
		Our aim here is to solve Eq.~\eqref{zeta}: $
		\rho	\big(\partial_t^2 \hat \zeta(\mathbf{k}, t) + \omega(k)^2 \hat \zeta(\mathbf{k}, t)\big) = - k \hat P_{\text{e}}(\mathbf{k}) \text{e}^{- \text{i} \mathbf{k} \bcdot \mathbf{r_0}(t)}
$,
		with the initial conditions \eqref{c_1} and \eqref{c_2}.
	     One readily notices that we have an explicit basis of the solution space at our disposal:
$y_1(t) = \text{e}^{\text{i} \omega(k) t}$ and  $y_2(t) = \text{e}^{-\text{i} \omega(k) t}$.
		One can therefore look for solutions of the form:
		\begin{equation}
			\hat \zeta(\mathbf{k}, t) = \lambda(t) y_1(t) + \mu(t) y_2(t)\, ,
		\end{equation}
		$\lambda$ and $\mu$ being two smooth functions of class $\mathcal{C}^1$ which depend on $\mathbf{k}$ implicitly. One can find the exact solution under the prescribed initial conditions by solving the system:
		\begin{equation}
			\begin{pmatrix}
				y_1 & y_2 \\
				y_1' & y_2'
			\end{pmatrix}
			\begin{pmatrix}
				\lambda' \\
				\mu'
				\end{pmatrix}
			=
			\begin{pmatrix}
				0 \\
				-\frac{k \hat P_\text{ext}(\mathbf{k}) \text{e}^{-\text{i} \mathbf{k} \bcdot \mathbf{r_0}(t)}}{\rho}
			\end{pmatrix},
		\end{equation}			
		Upon identification and integration, one is left with:
		\begin{eqnarray}
			\hat \zeta(\mathbf{k}, t) &=& \left(c(\mathbf{k}) - \frac{k}{2 \text{i} \omega(k) \rho} \int_0^t \text{e}^{-\text{i} \omega(k) \tau} \hat P_\text{e}(\mathbf{k}, \tau) \;\text{d} \tau \right)\text{e}^{\text{i} \omega(k) t} + \left(d(\mathbf{k}) + \frac{k}{2 \text{i} \omega(k) \rho} \int_0^t \text{e}^{\text{i} \omega(k) \tau} \hat P_\text{e}(\mathbf{k}, \tau) \;\text{d}\tau\right)\text{e}^{-\text{i} \omega(k) t}.
		\end{eqnarray}
		The initial conditions \eqref{c_1} and \eqref{c_2} yield that $c(\mathbf{k}) = d(\mathbf{k}) = 0$, and consequently  Eq.~\eqref{general_equation}.
	
	\section{The case of sudden uniform linear motion}\label{computation_zeta_hat}
		In the case of a sudden uniform linear motion, the expression of $\hat{\zeta}$ reduces to:
		\begin{equation}
			\hat \zeta(\mathbf{k}, t) = - \int_0^t \sin(\omega(k)(t - \tau)) \frac{k \hat P_\text{e}(\mathbf{k}) \text{e}^{-\text{i} v k_x \tau}}{\rho \omega(k)} \;\text{d} \tau.
		\end{equation}				
		Working out the integral yields:
		\begin{equation}
			\hat \zeta(\mathbf{k}, t) = -\frac{k \hat P_\text{e}(\mathbf{k})}{\rho \omega(k)} \frac{1}{\omega(k)^2 - (v k_x)^2} \left[\text{e}^{-\text{i} v k_x t} \omega(k) - \cos(\omega(k) t) \omega(k) + \text{i} \sin(\omega(k) t) v k_x \right].
		\end{equation}
		Now, with $\zeta_\text{d}(\mathbf{r}, t)$ being the surface elevation in the ship's frame of reference, we know that 
			$\zeta_\text{d}(\mathbf{r}, t) = \zeta(\mathbf{r} + v t \mathbf{u_x}, t)$, 
		and thus:
		\begin{equation}
			\hat \zeta_\text{d}(\mathbf{k}, t) = \hat \zeta(\mathbf{k}, t) \text{e}^{\text{i} v k_x t},
		\end{equation}
		which yields the result.
		
        \section{Proof of convergence of the wake for linear motion}\label{convergence}
	    We first show that the following term in equation~\eqref{zeta_hat_expression} vanishes as $t \to +\infty$ for a sudden uniform linear motion:
		\begin{equation}
			\zeta_{\text{d}, 2}(\mathbf{r}, t) = \frac{1}{(2 \pi)^2} \iint \text{e}^{\text{i} \mathbf{k} \bcdot \mathbf{r}} \frac{k \hat P_\text{e}(\mathbf{k})}{\rho (\omega(k)^2 - (v k_x)^2)} \text{e}^{\text{i}(v k_x + \omega(k)) t} \;\text{d}^2 \mathbf{k}\, .
		\end{equation}
		A common trick (see e.g.~\cite{richard1999capillary}) to overcome the indefiniteness 
	of the integral above, due to poles sitting on the integration domain, is to take into account the weak viscosity of the fluid $\nu$, which changes the denominator to $\rho (\omega(k)^2 - (v k_x)^2)$ to $\rho (\omega(k)^2 - (v k_x)^2 + i \pi \nu)$ in the equation above, by that shifting the poles of the integrated function off the real axis, and finally take the limit $\nu \to 0$. For the sake of simplicity, we leave such considerations out of the picture.
		Changing the variables to polar coordinates $(r, \varphi)$ yields:
		\begin{eqnarray}
			&&\zeta_{\text{d},2}(\mathbf{r}, t) = \frac{1}{(2 \pi)^2} \int_{\theta = 0}^{2 \pi} I(\theta, t) \;\text{d} \theta \, ,\nonumber \\
			&&\text{with} \quad I(\theta, t) = \int_{k = 0}^{+\infty} \frac{k \hat P_\text{e}(\mathbf{k})}{\rho(\omega(k)^2 - (v k \cos(\theta))^2)} \text{e}^{\text{i} k r \cos(\theta - \varphi)} \text{e}^{\text{i} (v k \cos(\theta) + \omega(k)) t} \;\text{d} k\, .
		\end{eqnarray}
		Let $f_\theta(k) = v k \cos(\theta) + \omega(k)$.
		If $\cos(\theta) \geq 0$, then the Riemann--Lebesgue lemma gives that $I(\theta, t) \to 0$ when $t \to+\infty$.
		Whenever $\cos(\theta) < 0$, $f_\theta$ has a unique critical point, that we name $k^\theta$. We then use the stationary phase approximation, which gives that $I(\theta, t) = ({C}/{\sqrt t}) \text{exp}({\text{i} t f_\theta(k^\theta)}) + \mathcal{O}\left({1}/{t}\right)$, where $C$ is a constant factor involving, among others, ${f_\theta}''(k^\theta)$. Notably, we still have $I(\theta, t) \to 0$.
		We conclude using Lebesgue's {Dominated Convergence Theorem}.\\
				
		We now prove that in the case of any linear motion reaching a final speed $v_0$ at a finite time $t_0$, the wake pattern in the ship's frame of reference converges to a constant as $t \to + \infty$.
		We note $r_0(\tau) = v_0 \tau - D$ for $\tau \geq t_0$. As before, we have $\hat \zeta_\text{d}(\mathbf{k}, t) = \hat \zeta(\mathbf{k}, t) \text{e}^{\text{i} v_0 k_x t}$, which enables us to write: 
		\begin{eqnarray}
			\hat \zeta_\text{d}(\mathbf{k}, t)=- \frac{k \hat P_\text{e}(\mathbf{k})}{\rho \omega(k)} \Bigg[ \int_0^{t_0} \sin(\omega(k) (t - \tau)) \text{e}^{\text{i} k_x (v_0 t - r_0(\tau))} \text d\tau &+& \text{e}^{\text{i} k_x D} \int_0^t \sin(\omega(k) (t - \tau)) \text{e}^{\text{i} k_x v_0 (t - \tau)} \text d\tau   \nonumber \\
			 &-& \text{e}^{\text{i} k_x D} \int_0^{t_0} \sin(\omega(k) (t - \tau)) \text{e}^{\text{i} k_x v_0 (t - \tau)}\text d\tau \Bigg].
		\end{eqnarray}
		The first and the third term can be treated like $\zeta_{\text{d}, 2}$ above, using the stationary phase approximation. As for the second term, in the Fourier domain, it is none other than  $\hat \zeta_\text{d}$, as given in Eq.~\eqref{zeta_hat_expression} (and broke down into five terms), with a factor $\text{e}^{\text{i} k_x D}$. We have already shown that the corresponding term in the real domain tends to a constant pattern in the ship's frame of reference, hence the result.

\end{document}